\documentstyle[11pt,newpasp,twoside,epsf]{article}
\def\edcomment#1{\iffalse\marginpar{\raggedright\sl#1\/}\else\relax\fi}
\reversemarginpar

\begin{document}
\title{The VLA Galactic Plane Survey (VGPS) }
 \author{Felix J. Lockman}
\affil{National Radio Astronomy Observatory, 
Green Bank, WV}
\author{Jeroen M.  Stil}
\affil{Department of Physics and Astronomy, 
University of Calgary}

\begin{abstract}
The VLA Galactic Plane Survey (VGPS) consists of  measurements of the 
21cm HI line and 1420 MHz continuum at $1\arcmin$ angular resolution 
over much of the first quadrant of Galactic longitude 
 within a few degrees of the Galactic plane.
 In combination with similar surveys 
of the fourth longitude quadrant and the outer Galaxy 
made with other instruments, 
 the VGPS will provide comprehensive data on the 
interstellar medium in most of the regions of active star formation in 
the Galaxy.  This paper describes the parameters of the VGPS, 
reports on its status, 
and discusses some early scientific results from the survey.

\end{abstract}

\section{Introduction }

Following the success of the Canadian Galactic Plane Survey 
(CGPS) of the 
northern Milky Way (Taylor et al 2003), work was begun on 
two additional surveys: 
 the Southern Galactic Plane Survey of the 
fourth longitude quadrant made with the Australia Telescope 
Compact Array (McClure-Griffiths 2002), 
and the VLA survey of the first quadrant of Galactic longitude 
(Taylor et al 2002).  
The goal of all these surveys is to provide a comprehensive, 
uniform data set in the 21cm HI line and 
1420 MHz continuum at an angular resolution of $1\arcmin$ covering 
the sky  within a few degrees  of the Galactic plane.  With such 
a data set it would be possible to make 
 synoptic studies of the interstellar medium  throughout the 
Galaxy.  Initial results from the surveys  show that the 
combination of high angular resolution and large area coverage 
reveals interstellar structures which were not known previously 
(e.g., Normandeau, Taylor, \& Dewdney 1996;
Gray et al 1998; Knee \& Brunt 2001; McClure-Griffiths et al 2000, 2002).  
As the portion of the Galaxy covered by the 
VLA observations contains important  regions of star formation and 
many molecular clouds, it is expected that this will be a particularly 
rich data set.

\section{Description of the VGPS}

\begin{figure}
\plotfiddle{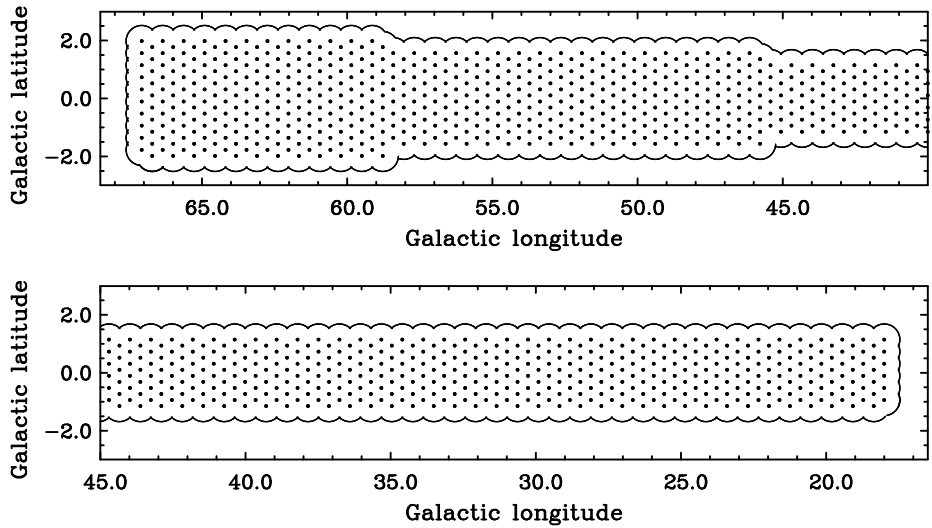}{3in}{-90}{50}{50}{-200}{240}
\caption{Pointing positions for the VLA Galactic Plane Survey (dots). 
The outline shows the resulting area covered to the half-power point
of the VLA primary beam.}
\end{figure}

 The area covered by the VLA Galactic Plane Survey (VGPS) 
 is shown in Figure 1.  Over this 
part of the sky the  VLA in D-configuration was used to observe 
 990 positions on a grid spacing of  $25'$.  
The pointing positions lie within Galactic latitudes $|b| \leq 1\fdg2$ 
at the lower longitudes and $|b| \leq  2\fdg0$ at the 
higher longitudes.   An important 
feature of the surveys is the addition of zero-spacing data to 
provide  the information on large angular scale emission which 
is missed by the interferometer.  
  Zero-spacing data for the 1400 MHz continuum were obtained 
from observations with the Effelsberg 100-meter telescope
 (Reich \& Reich 1986; Reich, Reich \& F\"urst 1990), while 
zero-spacing data for the 21cm HI line were supplied from 
 a brief survey of the entire area  made with 
 the new 100-meter  Robert C. Byrd Green Bank Telescope (GBT).  

At the VLA, HI spectra were taken over 256 channels separated by  
1.28 km s$^{-1}$, 
centered on a velocity which varied with longitude to track Galactic 
rotation.  The spectra from the two senses of circular polarization were 
offset by about 64 km s$^{-1}$ and one-half channel 
from each other to provide coverage of velocities 
 beyond that expected for 
Galactic HI. 
This produced an ample number of channels on either end of 
each spectrum which did not contain HI emission and 
could therefore be used to determine the  continuum level.  
 The result is coverage at full sensitivity of about 
240 km s$^{-1}$ of the 21cm spectrum.

\subsection{Observations}

At the VLA 
each field was observed in several $\sim$3 minute snapshots at 
different hour angles to  vary the UV coverage.  The field 
centers were separated by $25\arcmin$, which is 80\% of the primary VLA
beam FWHM.  The GBT data  have $9\arcmin$ angular resolution and were taken 
in a quick raster scan across the Galactic plane.  
Standard gain and phase calibration was performed for the VLA data, but the 
 flux scale for these data is calibrated with respect to fluxes within each pointing of sources 
observed in the NVSS (Condon et al 1998).  
In the final data processing the 
GBT flux scale is forced into agreement with that of the VLA, on 
average, for baselines  which overlap in the two data sets.

\begin{table}
\begin{center}
\caption{Parameters of the VLA Galactic Plane Survey}
\begin{tabular}{ll}
\tableline 
Longitude & $18\deg \leq \ell \leq 67\deg$ \\
Latitude & $\pm 1\fdg2 \rightarrow \pm2\deg$ \\
Angular Resolution & $1\farcm0$ \\
Vel. Coverage & 240 km s$^{-1}$ \\
Vel. Resolution & 1.5 km s$^{-1}$ \\
Noise (Continuum) ($1\sigma$) & 2 mJy per beam \\
Noise (HI) ($1\sigma$) & 2.0 K \\
\tableline
\tableline \\
\end{tabular}
\end{center}
\end{table}

\subsection{Image Processing}

Continuum subtraction was done in the UV plane by  fitting a linear
function to the visibilities in line-free channels.  
Strong, compact continuum sources were cleaned by using data only 
from  projected
baselines longer than $0.3\ \rm k\lambda$.  These absorbed sources
were subtracted from the  HI images and restored with a Gaussian
beam before deconvolution of the line emission. This procedure
eliminates the sidelobes around absorbed continuum sources, in
particular, the radial spokes which occur on the 20\% level in the dirty
beam of VLA snapshots.  The HI line emission was then deconvolved with
an \ $I {\rm log} I$\ maximum entropy algorithm (Cornwell \& Evans 1985).  
 The deconvolved model of clean components was restored
with a $60\arcsec$ circular Gaussian beam which is 
uniform over the VGPS survey area.

Figure 2 gives an example of 
 the quality of the survey data and also illustrates the 
importance of the single-dish, low spatial-frequency data  
in deriving  complete information on Galactic HI.  
  Although the left section of the figure, which contains 
VLA data alone, shows considerable detail which is invisible 
in the lower-resolution GBT data  (right segment), it is 
only with combination of the two that both the detail and the 
correct overall amplitude of the HI is recovered.  The central 
section of Figure 2 is representative of the quality expected in
 one velocity channel of  the 
final, fully-processed HI  data. 

\begin{figure}
\plotfiddle{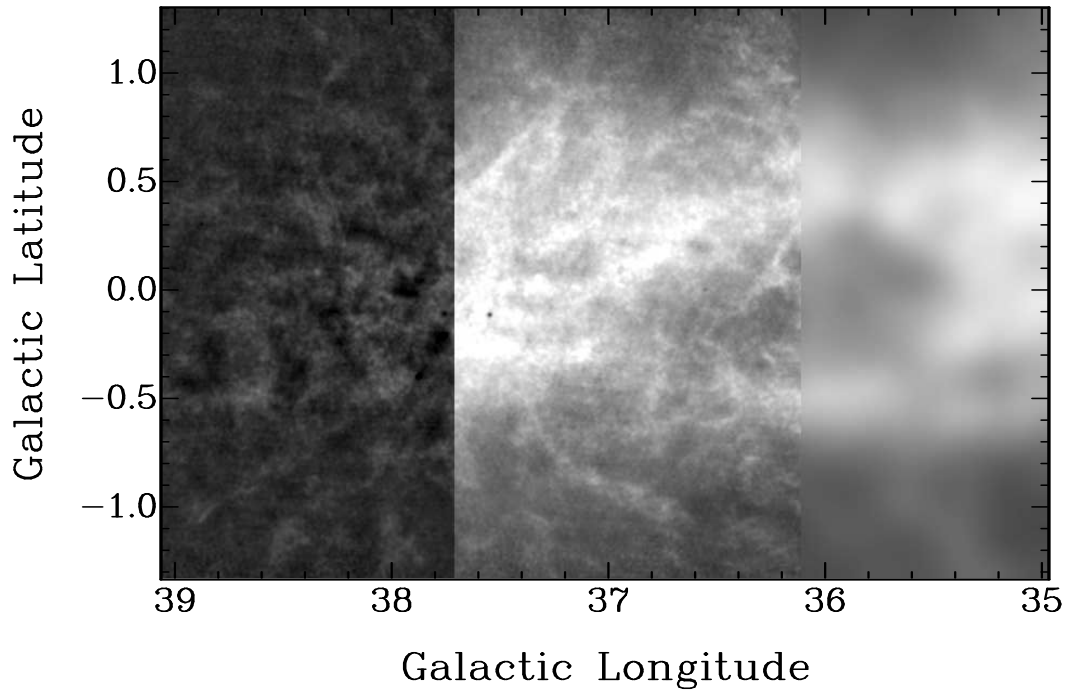}{2.5in}{90}{65}{65}{242}{-45}
\caption{The HI line at 90.5 km s$^{-1}$ in the Galactic Plane 
between $35\deg$ and $39\deg$. Longitudes in  the {\bf right} third 
of the figure contain  GBT data alone at $9\arcmin$ angular resolution.  
Longitudes in the {\bf left} third of the figure show only 
VLA data at $1\arcmin$ resolution.  Longitudes in the {\bf center} 
third of the Figure show the combined VLA and GBT data.  
The intensity scale is the same for all three images --- the VLA data look 
much dimmer because most of the power in Galactic HI is on angular 
scales to which the interferometer has no sensitivity.  
Only the central portion 
of the figure, which contains data from all spatial frequencies, has 
both the high angular resolution and full information.  This will be
typical of the final VGPS data.}
\end{figure}

\subsection{The Final Data Products}

The final VGPS data product will consist of continuum images and HI 21-cm
spectral line cubes with velocity channels 
spaced by 0.82 $\ \rm km\ s^{-1}$ (to match 
the velocity channels of the CGPS), angular resolution of $1\arcmin$, 
and r.m.s. sensitivity of 
2 K per velocity channel ($S_{\nu}/T_b = 5.94\
\rm mJy/K$).  At the time of this writing all the data have been 
edited and the final calibration of the 
individual VLA pointings is in progress.  Completed cubes should be
available for portions of the VGPS area by the end of 2003 and the 
complete data set will be  distributed in mid-2004.  The VGPS 
parameters are summarized in Table 1.

\section{Some Early Results}

\begin{figure}
\plotfiddle{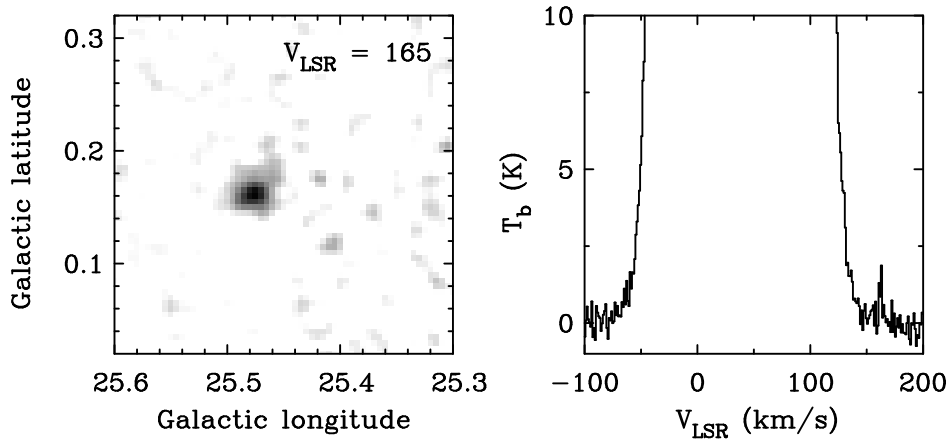}{2.5in}{-90}{50}{50}{-212}{282}
\caption{Left: VLA observations of the HI emission around longitude $25\fdg5$ 
 showing a compact, slightly-resolved HI cloud  at 165  km s$^{-1}$, 
nearly 50 km s$^{-1}$ past 
the maximum permitted by Galactic rotation.  The GBT HI spectrum 
showing the cloud is on the right.  This cloud may be 
related to the clouds discovered recently in GBT observations of 
the Galactic halo.  They  belong to a population which has a 
large cloud-cloud velocity dispersion. }
\end{figure} 

The survey data have already revealed several interesting HI features.
One is a well-defined, expanding  HI shell with $\approx 10^3 M_{\sun}$ 
in neutral gas found lying about 100 pc below the Galactic plane 
(Stil et al 2003).  Another is the remarkable cloud shown in Figure 3.  

This HI cloud, at $\ell,b,V = 25\fdg48,+0\fdg16,+165$ km s$^{-1}$  
has a velocity about 
50 km s$^{-1}$ greater than permitted by Galactic rotation in this 
direction, so one might initially suspect that it is some sort of compact 
 high-velocity HI cloud not associated with the Galactic disk.  
However, recent studies of HI above the Galactic plane using the 
GBT have shown that there is a population of HI clouds similar to this 
one, with velocities  determined mainly 
by circular Galactic rotation, but with a large cloud-cloud 
velocity dispersion of several tens of km s$^{-1}$ (Lockman 2002).  
These  ``halo'' HI clouds can be traced down to fairly low latitude 
before they blend with other HI emission and are lost to view.  
It is likely that G25.48+0.16 is a member of this group.  
Because of its very high random velocity it   can be 
 distinguished from confusing HI even near the plane.  
Similar clouds, though with a smaller random component of velocity,
    are seen elsewhere in the GBT data at low latitudes. 
The association with the halo clouds implies that G25.48+0.16 
most likely 
lies near the tangent point at a distance of 7.7 kpc, otherwise
its random velocity would be even more extreme than it already is.  

  The cloud's physical properties (diameter $\approx 5$ pc, $N_{HI} \approx 
2 \times 10^{20}$ cm$^{-2}$, $\langle n \rangle \approx 14$ 
cm$^{-3}$),  are not exceptional, and in fact are quite a good 
match to the McKee \& Ostriker (1977) model for a cold cloud 
(see their Table 1).  
But the 50 km s$^{-1}$  random velocity of the cloud 
 is perplexing, for 
if the cloud is located in a medium of $\langle n \rangle \sim 0.3$ 
cm$^{-3}$, it will sweep up its mass in only 10$^6$ years 
and thus be decelerated rapidly.  Most of the halo HI clouds 
are sufficiently far from the plane that their lifetime is not 
a major problem, but how can this 
cloud exist  at all, unless it is traveling through a region of 
very low density?   This cloud may be one of a population of 
 ``shocked interstellar clouds'' proposed by Radhakrishnan \& 
Srinivasan (1980; see also Kulkarni \& Fich 1985) to explain 
the extensive wings of HI profiles at low latitude.

These tantalizing 
results from a crude perusal of partially-reduced data give a taste 
of what is to come.  When the VGPS is available in its final form we 
expect that it will have a major impact on Galactic studies.  

\acknowledgements
We thank J. M. Dickey, Y. Pidoprihora and A.R. Taylor for helpful comments.
The National Radio Astronomy Observatory is operated by Associated 
Universities, Inc., under a cooperative agreement with the National 
Science Foundation.


\begin{references}

\reference 
Condon, J. J., Cotton, W. D., Greisen, E. W., Yin, Q. F., Perley, R. A., 
Taylor, G. B., \& Broderick, J. J. 1998, AJ, 115, 1693

\reference
Cornwell, T. J., \& Evans, K. F., 1985, A\&A 143, 77

\reference
Gray, A. D., Landecker, T.L., Dewdney, P.E., 
\& Taylor, A.R. 1998, Nature, 393, 660


\reference 
Knee, L.B.G. \& Brunt, C.M. 2001, Nature, 412, 308

\reference 
Kulkarni, S.R., \& Fich, M. 1985, ApJ, 289, 792

\reference 
Lockman, F.J. 2002, ApJ, 580, L47

\reference 
McClure-Griffiths, N.M.  2002, in  ``Seeing Through the Dust'', ASP Conf. Ser. 
Vol. 276, ed. A.R. Taylor, T.L. Landecker, \& A.G. Willis, p. 58

\reference 
McClure-Griffiths, N.M., Dickey, J.M., Gaensler, B.M.,  Green, A.J., 
Haynes, R.F., \& Wieringa, M.H. 2000, AJ, 119, 2828

\reference 
McClure-Griffiths, N.M., Dickey, J.M., Gaensler, B.M., \& 
Greene, A.J. 2002, ApJ, 578, 176

McKee, C.F., \& Ostriker, J.P. 1977, ApJ, 218, 148

\reference
Normandeau, M., Taylor, A.R., \& Dewdney, P.E. 1996, Nature, 380, 687

\reference 
Radhakrishnan, V., \& Srinivasan, G. 1980, J. Astr. Ap., 1, 47

\reference
 Reich, P., \& Reich, W., 1986, A\&AS, 63, 205

\reference
 Reich, W., Reich, P., \& F\"urst 1990, A\&AS, 83, 539 

\reference
Stil, J.M., Taylor, A.R., Martin, P.G., Rothwell, T., Dickey, J.M., \& 
McClure-Griffiths, N.M. 2003, in press

\reference
Taylor A.R. et al 2002, in ``Seeing Through the Dust'', ASP Conf. Ser. 
Vol. 276, ed. A.R. Taylor, T.L. Landecker, \& A.G. Willis, p. 68

\reference
Taylor, A.R. et al 2003, AJ, 125, 3145

\end{references}
\end{document}